\documentclass{aa}
\usepackage{graphicx,natbib}
\bibpunct{(}{)}{;}{a}{}{,}

\newcommand{\nh}{N_\mathrm{H}}

\newcommand{\lesssim}{\mathrel{\hbox{\rlap{\hbox{\lower4pt\hbox{$\sim$}}}\hbox{$<$}}}}
\newcommand{\gtrsim}{\mathrel{\hbox{\rlap{\hbox{\lower4pt\hbox{$\sim$}}}\hbox{$>$}}}}
\newcommand{\beq}{\begin{equation}}
\newcommand{\eeq}{\end{equation}}
\newcommand{\beqa}{\begin{eqnarray}}
\newcommand{\eeqa}{\end{eqnarray}}

\begin{document}

\title{Identification of 8 INTEGRAL hard X-ray sources with Chandra} 

\author{S.~Sazonov\inst{1,2} \and  E.~Churazov\inst{1,2} \and
 M.~Revnivtsev\inst{1,2} \and A.~Vikhlinin\inst{3,2} \and R.~Sunyaev\inst{1,2}}

\offprints{sazonov@mpa-garching.mpg.de}

\institute{Max-Planck-Institut f\"ur Astrophysik,
           Karl-Schwarzschild-Str. 1, D-85740 Garching bei M\"unchen,
           Germany
     \and   
           Space Research Institute, Russian Academy of Sciences,
           Profsoyuznaya 84/32, 117997 Moscow, Russia
     \and
           Harvard-Smithsonian Center for Astrophysics, 60 Garden St.,
           Cambridge, MA 02138, USA 
}
\date{Received / Accepted}

\authorrunning{Sazonov et al.}
\titlerunning{Identification of hard X-ray sources}

\abstract{We report the results of identification of 8 hard X-ray sources
discovered by the INTEGRAL observatory during the ongoing all-sky
survey. These sources have been observed by Chandra. In 6 cases a 
bright X-ray source was found within the INTEGRAL localization region, which
permitted to unambigously identify 5 of the objects with nearby
galaxies, implying that they have an active galactic nucleus (AGN),
whereas one source is likely an X-ray binary in LMC. 4 of the 5 newly
discovered AGNs have measured redshifts in the range 0.025--0.055. The
X-ray spectra reveal the presence of significant amounts of absorbing
gas ($\nh$ in the range $10^{22}$--$10^{24}$~cm$^{-2}$) in all 5 AGNs,
demonstrating that INTEGRAL is starting to fill in the sample of
nearby obscured AGNs.  

}
\maketitle

\section{Introduction}

X-ray surveys have been playing a leading role in studying the
demography and cosmological evolution of active galactic nuclei
(AGNs). In particular, deep surveys performed in the standard X-ray band
(2--10~keV) have dramatically improved our knowledge of AGNs at $z\gtrsim 0.3$
(e.g. \citealt{bh05}). However, due to the small solid angle, deep surveys
are not well suited for studying the local AGN population. Furthermore, there
is sufficient evidence that most AGNs are heavily intrinsically obscured at
optical and soft X-ray wavelengths. To obtain a full picture of the  
local AGN population there is therefore a need in very large area
surveys performed at energies above 10~keV. Such surveys should also
be useful for population studies of Galactic hard X-ray sources.  

Recently, a high Galactic latitude ($|b|>10^\circ$) survey of the whole sky in
the 3--20~keV band was performed with the RXTE observatory
\citep{rsj+04,sr04}, detecting in total $\sim 300$ sources, including $\sim 
100$ AGNs. Even this survey was however strongly biased against sources with
intrinsic absorption in excess of a few $10^{23}$~H atoms per cm$^2$. 

A comparable or better (for strongly absorbed sources and/or in crouded
regions) capability for studying bright ($\gtrsim 1$~mCrab) hard X-ray
sources is now provided (e.g. \citealt{kvc+05}) by the IBIS telescope
of the INTEGRAL observatory, owing to its sensitivity peaking above
20~keV, large field of view and good ($\sim$$10^\prime$) angular
resolution. INTEGRAL has been observing multiple targets since October
2002 and more than half of the sky has been 
covered as of spring 2005. A recently started campaign to observe with
INTEGRAL the remaining sky regions is to be completed in
2006. Once the survey of the whole sky is completed, it will
be used for a statistical study of extragalactic hard X-ray
sources. This will include the construction of the source number-flux
function as well as the hard X-ray luminosity function and
absorption column density distribution of local AGNs.

Although most of the INTEGRAL sources detected so far away from the
Galactic plane and Galactic Center can be reliably associated with
known AGNs, there is a significant fraction of unidentifed 
sources. We hope to identify most of these sources by improving their 
localization to few arcseconds using focusing X-ray telescope
observations, so that optical identification then becomes
straightforward. In the present pilot study we use Chandra to identify
8 hard X-ray sources discovered by INTEGRAL. 

\section{Observations}

Public and our proprietary INTEGRAL/IBIS observations cover approximately
50\% of the sky as of April 2005. We constructed a map of the observed
sky in the 17--60~keV band from the data of the IBIS/ISGRI
detector. The details of the analysis and the catalog of
detected sources will be presented elsewhere, after new observations
from the ongoing all-sky survey are added and analyzed.
 
\begin{table*}[htb]
\caption{Identification of INTEGRAL sources
\label{ident_table}
}
\smallskip

\begin{tabular}{lrcrll}
\hline
\hline
\multicolumn{3}{c}{INTEGRAL} &
\multicolumn{1}{c}{Chandra} &
\multicolumn{1}{c}{Identification} &
\multicolumn{1}{c}{Class of object}
\\
\cline{1-3}
\multicolumn{1}{c}{Name} &
\multicolumn{1}{c}{$\alpha$, $\delta$ (2000)} &
\multicolumn{1}{c}{Err} &
\multicolumn{1}{c}{$\alpha$, $\delta$ (2000)} &
\multicolumn{1}{c}{} &
\multicolumn{1}{c}{}
\\
\hline
IGR J05007$-$7047 &  75.217 $-$70.771 &  3' &   05:00:46.08 $-$70:44:36.0 &
USNO-B1.0 0192-0057570  & HMXB in LMC? \\
IGR J07563$-$4137 & 119.070 $-$41.635 &  3'  &   07:56:19.62 
$-$41:37:42.1&
2MASX J07561963$-$4137420 & AGN \\
IGR J10252$-$6829 & 156.267 $-$68.483 &  3'  &   10:25:00.49 $-$68:27:27.3&
                           & \\
IGR J11085$-$5100 & 167.144 $-$51.014 &  3'  &   11:08:50.48 $-$51:02:32.8&
USNO-B1.0 0389-0243808?   & \\
IGR J12026$-$5349 & 180.657 $-$53.816 &  3'  &   12:02:47.63 $-$53:50:07.7&
WKK 0560                  & AGN (z=0.0280)\\
IGR J12391$-$1612 & 189.780 $-$16.196 &  2'  &   12:39:06.29 $-$16:10:47.1 &
2MASX J12390630$-$1610472 & AGN (z=0.0367)\\
IGR J13091+1137   & 197.267   +11.622 &  3'  &   13:09:05.60   +11:38:02.9 &
NGC 4992                  & AGN (z=0.0251)\\
IGR J19473+4452   & 296.836   +44.864 &  3'  &   19:47:19.37   +44:49:42.4&
2MASX J19471938+4449425   & AGN (z=0.0539)   \\
\hline
\end{tabular}

\end{table*}

For the present study we compiled a representative sample
of 8 new hard X-ray sources, detected with statistical significance
higher than 5.5$\sigma$ on the summed INTEGRAL map. These sources are
located significantly far away from both the Galactic plane ($>$$5^\circ$)
and Galactic Center ($>$$60^\circ$), and were not detected in the
Rosat All-Sky Survey
(http://www.xray.mpe.mpg.de/cgi-bin/rosat/rosat-survey). The ISGRI  
localizations are 2--3' radius (90\% confidence). This INTEGRAL sample  
was observed by Chandra/ACIS-I as part of the DDT program
in June--July 2005, with an exposure of 3.5~ks per source.

\subsection{Identification}

\begin{figure*}
\centering
\includegraphics[width=0.72\textwidth]{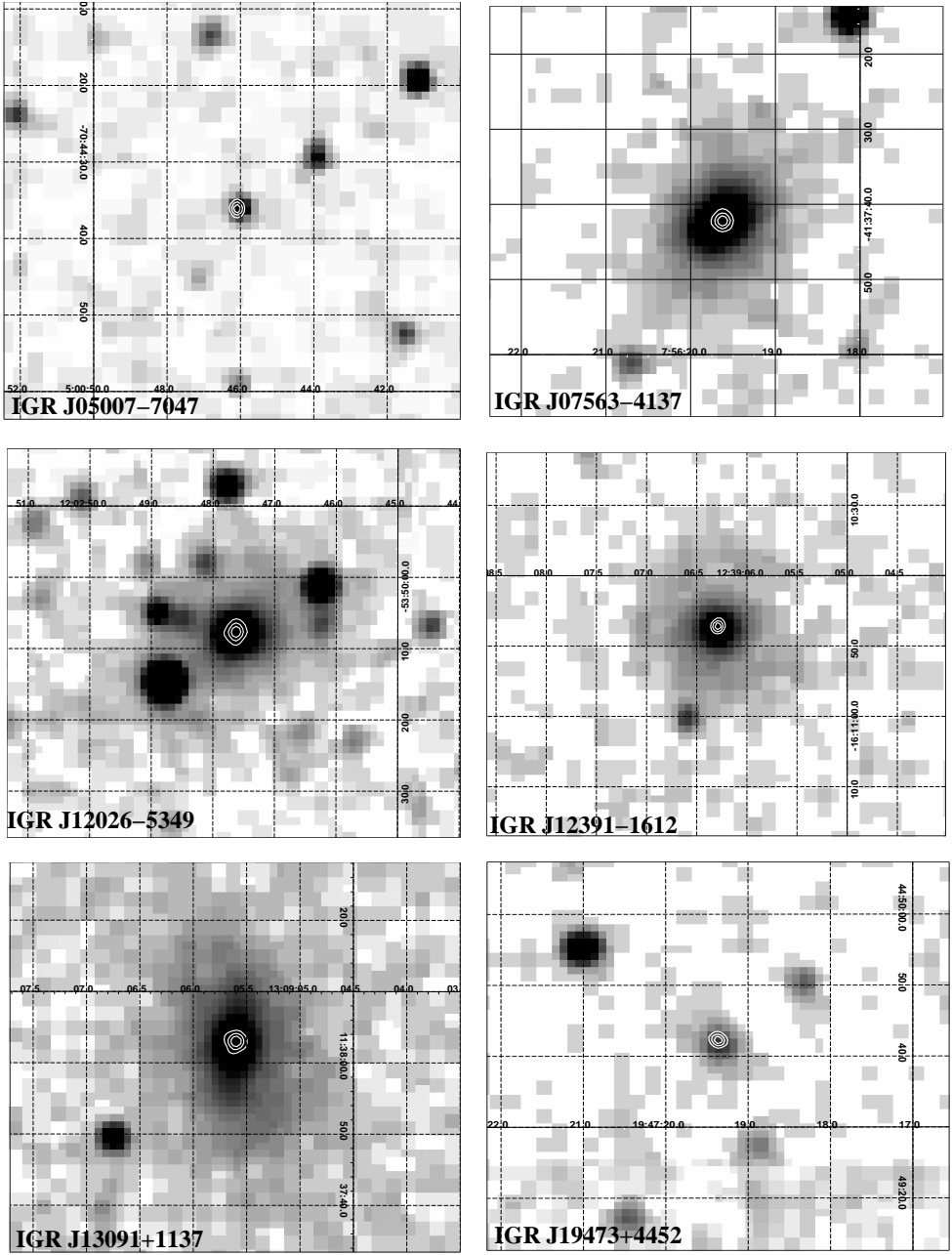}
\caption{Chandra X-ray brightness contours overlayed on the 2MASS
J-band images of 6 INTEGRAL sources. Each image ($\sim
0.8^\prime\times 0.8^\prime$) falls into the original INTEGRAL
localization region ($\sim 3^\prime$ in radius, not shown). 
}
\label{images_fig}
\end{figure*}

In the Chandra fields for 6 INTEGRAL sources there is
a bright point source ($\sim$0.1--0.2 cts~s$^{-1}$) inside the
ISGRI error region that can be unambiguously associated 
with the IGR source, since sources with corresponding X-ray fluxes
(a few $10^{-12}~{\rm erg}~{\rm cm}^{-2}~{\rm s}^{-1}$) are
quite rare ($\lesssim 1$ per 10 sq.~deq.) at high Galactic latitude
(e.g. \citealt{rsj+04}). In Fig.~\ref{images_fig} the Chandra
detections are shown on the near-infrared images from the Two
Micron All-Sky Survey. Search  around the accurate ($\sim$0.6'')
Chandra positions in NED and SIMBAD reveals that 5 of  
the bright X-ray counterparts are located in the central regions of nearby
galaxies, which together with the inferred high X-ray 
luminosities (see Table~\ref{spec_table}) implies that they are
AGNs. For 3 of these galaxies there is a published redshift. We have
measured the redshift of the 4th AGN, 2MASX J19471938+4449425, with
the RTT-150 telescope \citep{b05}. The distance to 2MASX
J07561963-4137420 remains unknown.   

The position of the bright Chandra counterpart to IGR J05007$-$7047 
coincides with the relatively bright, blue ($V=14.8$, $B-V=-0.01$,
\citealt{m02}) star USNO-B1.0 0192-0057570. Based on its magnitude, color and
positional coincidence with the Large Magellanic Cloud, we conclude that this
object is very likely a high-mass X-ray binary in LMC. With the
luminosity $\sim 5\times 10^{36}$~erg~s$^{-1}$ (1--60~keV, assuming a distance
of 50~kpc), it is then the 4th brightest HMXB in LMC (after LMC X-1, LMC
X-3 and LMC X-4) according to INTEGRAL observations and historical
data (e.g. \citealt{sg05}). 
 
In either localization region of the 2 remaining sources,
IGR J10252$-$6829 and IGR J11085$-$5100, there is a single relatively bright
($\sim$0.003 and $\sim$0.005~cts~s$^{-1}$, respectively) Chandra source,
which might be associated with the IGR sources. However, the
probability of the Chandra source occuring by chance inside the
INTEGRAL error region is not small in both cases, as suggested by 
counts of extragalactic X-ray sources (e.g. \citealt{mcl+03}) and by
the fact that there are 5 (3) sources brighter than 
0.0015~cts~s$^{-1}$ in the entire ($\sim 16^\prime\times 16^\prime$)
Chandra field for IGR J10252$-$6829 (IGR J11085$-$5100). Nonetheless,
we provide the positions of the possible X-ray counterparts of both
IGR sources in Table~\ref{ident_table} and note that the possible
counterpart of IGR J11085$-$5100 coincides with a weak ($B=20.3$)
star-like object. New INTEGRAL data suggest that both IGR
J10252$-$6829 and IGR J11085$-$5100 may be transient sources 
since they were clearly detectable in 2004 but disappeared in 2005.  

\subsection{X-ray spectra}

\begin{figure}[htb]
\centering
\includegraphics[bb=20 170 430 720, width=\columnwidth]{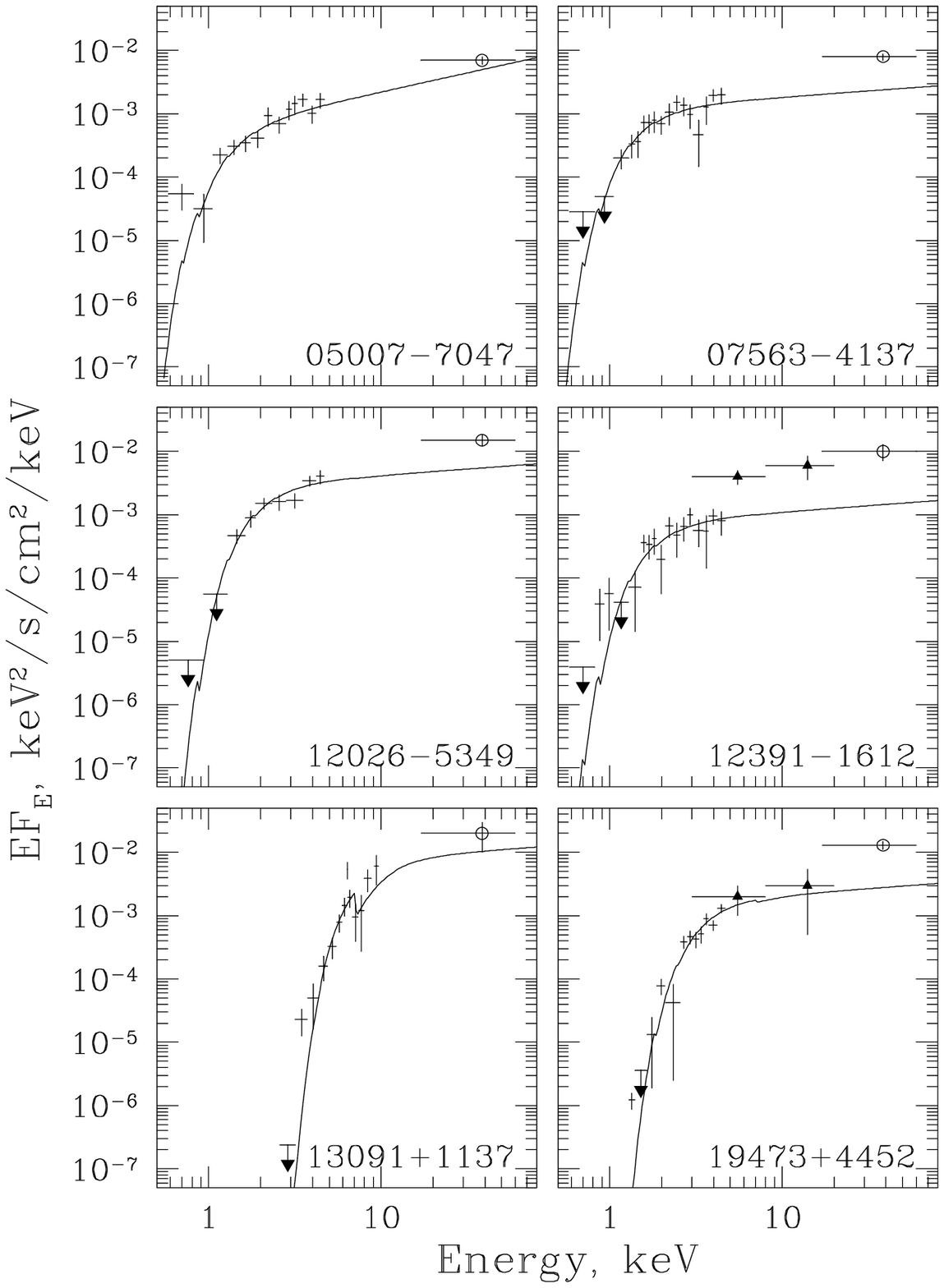}
\caption{Broad-band X-ray spectra of 6 IGR sources with bright Chandra
counterparts. The data points between 0.5 and 5~keV (0.5--10~keV for IGR
J13091+1137) are from Chandra and the solid line is the
best-fit model to these data (see Table~\ref{spec_table}) extrapolated
to higher energies. Also shown are 17--60~keV fluxes measured by
INTEGRAL (open circles) and, for 2 sources, 3--8~keV and 8--20~keV fluxes
measured by RXTE (filled triangles). 
}
\label{spectra_fig}
\end{figure}

We next discuss the X-ray spectra of the IGR sources. Chandra provides
information on the source spectra  below 10~keV, while INTEGRAL yields
the source fluxes in the 17--60~keV band. Two of the sources,
IGR J12391$-$1612 and IGR J19473+4452, were also detected in the RXTE Slew
Survey \citep{rsj+04} where their fluxes in the 3--8~keV and
8--20~keV bands were measured. We show in Fig.~\ref{spectra_fig} the
resulting broad-band spectra for the 5 AGNs and the candidate HMXB in LMC. 
  
The data for the 5 brightest Chandra sources suffer from photon pileup
(e.g. \citealt{d01}). A generic singature of the pileup in the
ACIS CCDs is photon migration from the good (0,2,3,4,6) to bad (1,5,7)
grade set (see \citealt{d01} for details). The fraction of bad grade
photons is negligible for non piled-up sources. Therefore, the image
regions strongly affected by pileup can be identified from the image
containing only the bad grade photons. For our IGR sources, the
affected regions are mostly confined to the central 1.5x1.5
arcsec. Outside these regions, the pileup effects must be weak 
because the fraction of bad grade photons is small ($<10$\%). We
based our spectral analysis for the piled-up sources on the source
photons detected outside the central regions (i.e. excluding
$\sim$80\% of the PSF) and also restricted the analysis to energies
below 5~keV. The nominal effective area calibration was corrected for
the energy dependence of the flux scattered outside the region
affected by pileup (http://cxc.harvard.edu/cal/Hrma/psf/). To construct
the spectrum of IGR~J13091+1137, unaffected by pileup, all photons
from the whole PSF were used. 

The Chandra spectra of all 5 AGNs are well approximated by a power 
law modified by photoabsorption along the line of sight. All 5 spectra
exhibit a strong low-energy cutoff due to intrinsic absorption, with
inferred column densities ($\nh$) ranging from $10^{22}$ to
$10^{24}$~cm$^{-2}$ (see Table~\ref{spec_table}). This explains the absence of 
soft X-ray counterparts in the Rosat All-Sky Survey. For all 5 spectra
the photon index ($\Gamma$) is constrained to the broad range 1.3--2.0.  
The best-fit models shown in Fig.~\ref{spectra_fig} assume $\Gamma=1.8$, a
value typical for AGNs. It can be seen that the extrapolated  best-fit
model misses the INTEGRAL and RXTE flux measurements by up to a factor
of 5. This discrepancy likely results from a combination of uncertainty in the
spectral slope and intrinsic source variability.

Similarly to the AGNs, the Chandra spectrum of IGR J05007$-$7047, the
proposed LMC HMXB, is well fit by a power law ($\Gamma$ in the
range 1.3--2.0) modified by photoabsorption ($\nh\approx
10^{22}$~cm$^{-2}$). In Fig.~\ref{spectra_fig} we 
fixed $\Gamma$ at 1.4, a value typical for HMXBs. The extrapolated best-fit
model is in good agreement with the 17--60~keV flux measured by
INTEGRAL.

\begin{table*}
\begin{center}
\caption{X-ray fluxes, luminosities and spectra
\label{spec_table}
}
\smallskip

\begin{tabular}{cccccccc}
\hline
\hline
\multicolumn{1}{c}{Source} &
\multicolumn{1}{c}{$F$ (0.5--8 keV)} &
\multicolumn{1}{c}{$L$ (0.5--8 keV)$^{\rm a}$} &
\multicolumn{1}{c}{$\Gamma$} &
\multicolumn{1}{c}{$\nh$$^{\rm b}$} &
\multicolumn{1}{c}{$N_\mathrm{H, Gal}$$^{\rm c}$} &
\multicolumn{1}{c}{$F$ (17--60 keV)} &
\multicolumn{1}{c}{$L$ (17--60 keV)$^{\rm a}$}
\\
\multicolumn{1}{c}{} &
\multicolumn{1}{c}{$10^{-12}$~erg/cm$^{2}$/s} &
\multicolumn{1}{c}{erg/s} &
\multicolumn{1}{c}{fixed} &
\multicolumn{2}{c}{$10^{22}$~cm$^{-2}$} &
\multicolumn{1}{c}{$10^{-12}$~erg/cm$^{2}$/s} &
\multicolumn{1}{c}{erg/s}
\\
\hline
IGR J05007$-$7047 & $3.0\pm 0.3$ & $(9.1\pm 0.9)\times 10^{35}$     &
 1.4 & $1.0\pm 0.2$ & 0.08 & $12\pm 2$ & $(3.6\pm 0.6)\times 10^{36}$ \\
IGR J07563$-$4137 & $3.5\pm 0.4$ &  ---                             &
 1.8 & $1.1\pm 0.2$ & 0.43 & $15\pm 2$ &   ---                \\
IGR J12026$-$5349 & $6.7\pm 0.9$ & $(1.1\pm 0.1)\times 10^{43}$     &
 1.8 & $2.2\pm 0.3$ & 0.17 & $33\pm 3$ & $(5.3\pm 0.5)\times 10^{43}$ \\
IGR J12391$-$1612 & $2.0\pm 0.3$ & $(5.5\pm 0.8)\times 10^{42}$ &
 1.8 & $1.9\pm 0.3$ & 0.04 & $42\pm 7$ & $(1.1\pm 0.2)\times 10^{44}$ \\
IGR J13091+1137   & $1.2\pm 0.2$ & $(1.5\pm 0.3)\times 10^{42}$     &
 1.8 & $90\pm 10$   & 0.02 & $34\pm 5$ & $(4.3\pm 0.6)\times 10^{43}$ \\ 
IGR J19473+4452   & $3.0\pm 1.0$ & $(1.8\pm 0.6)\times 10^{43}$     &
 1.8 & $11\pm 1$    & 0.17 & $25\pm 4$ & $(1.5\pm 0.2)\times 10^{44}$ \\  
\hline
\end{tabular}
\end{center}
$^{\rm a}$ Observed luminosities, assuming
$H_0=73$~km~s$^{-1}$~Mpc$^{-1}$ for AGNs and $D=50$~kpc for IGR
J05007$-$7047 

$^{\rm b}$ Measured absorption column 

$^{\rm c}$ Galactic absorption column based on \cite{dl90} 

\end{table*}

Table~\ref{spec_table} provides source fluxes at 0.5--8~keV and 17--60~keV as
well as luminosities if the distance is known. The found
AGN luminosities are typical for Seyfert galaxies.   

\section{Discussion}

We observed with Chandra 8 hard X-ray sources discovered by
INTEGRAL. 5 AGNs and a likely HMXB in LMC have been found. 
As expected, most of the new hard X-ray sources proved nearby
($z=0.025$--0.055 in the 4 cases with measured redshifts) AGNs with
significant intrinsic absorption. This verifies our expectation of
discovering a large number of obscured AGNs (mostly Seyfert~2 galaxies)
during the INTEGRAL all-sky survey. We note that INTEGRAL has also
demonstrated its capability of discovering Seyfert~1 galaxies hidden behind
the Milky Way (see e.g. \citealt{mpb+04}).

This study demonstrates that the majority of hard X-ray sources
discovered and localized by INTEGRAL at high Galactic latitude can be
readily identified using short Chandra, XMM-Newton or SWIFT/XRT exposures. 

\smallskip
\noindent {\sl Acknowledgments} This research is based on Chandra observations
using Chandra Director's Discretionary Time. We are grateful
to Harvey Tananbaum for his interest in the problem and useful
discussions, and to William Forman for help. This research has made
use of the SIMBAD database (operated at CDS, Strasbourg), NASA/IPAC
Extragalactic Database and the Infrared Science Archive (operated by
the Jet Propulsion Laboratory, California Institute of Technology),
and data products of the 2MASS (joint project of the Univ. of
Massachusetts and the IPAC, funded by the NASA and NSF).


\end{document}